\documentclass[a4paper]{jpconf}
\usepackage{graphicx}
\usepackage{amsmath, amssymb}
\begin{document}
\title{Photoinduced insulator-metal transition in correlated electrons 
--- a Floquet analysis with the dynamical mean-field theory}

\author{Naoto Tsuji$^1$, Takashi Oka and Hideo Aoki}

\address{Department of Physics, University of Tokyo, Hongo 7-3-1, 
Bunkyo-ku, Tokyo 113-0033, Japan}

\ead{$^1$ tsuji@cms.phys.s.u-tokyo.ac.jp}

\begin{abstract}
In order to investigate photoinduced insulator-metal transitions 
observed in correlated electron systems, 
we propose a new theoretical method, where we combine a Floquet-matrix method 
for AC-driven systems with the dynamical mean-field theory. The method can treat 
nonequilibrium steady states exactly beyond the linear-response regime. 
We have applied the method to the Falicov-Kimball model coupled 
to AC electric fields, and numerically obtained the spectral function, 
the nonequilibrium 
distribution function and the current-voltage characteristic.  
The results show 
that intense AC fields indeed drive Mott-like insulating states 
into photoinduced metallic states in a nonlinear way. 
\end{abstract}

\section{Introduction}

There is an increasing fascination with 
photoinduced phase transitions (PIPTs), where a 
`phase transition' is triggered by  a laser of 
strong intensity.  
Specifically, strongly correlated electron systems harbor 
the phenomena, where an insulating state in, {\it e.g.}, 
manganese oxides can be made 
metallic when a pulsed laser is injected \cite{Miyano, Fiebig}.  
In some cases a 
transient ferromagnetic spin alignment can be generated as observed in recent
pump-probe spectroscopy experiments \cite{Matsubara}.  
While these phenomena are intrinsically nonequilibrium, 
a remarkable feature is 
the threshold behavior, namely,  
the transition only occurs when the intensity of light exceeds a 
certain threshold value. This implies that PIPT is essentially beyond 
the linear-response regime for the external driving fields. 
It is then a theoretical challenge to identify 
the nature of the states emerging in nonequilibrium 
against the frequency $\Omega$ and the intensity $E$ of the AC field.

A crucial aspect in theoretically studying 
PIPT in strongly correlated systems is that we have to simultaneously 
take account of 
both the electron correlation 
and the nonlinear effect of the electric field. 
Here we propose a theoretical method, in which 
we incorporate the Floquet-matrix 
method for AC field problems \cite{Shirley, Sambe} into 
the nonequilibrium dynamical 
mean-field theory (DMFT) \cite{Freericks}.  
The Floquet approach can treat a 
nonequilibrium steady state nonperturbatively ({\it i.e.} up to 
arbitrary orders of $E$), while the DMFT can describe correlation 
effects like Mott's transition. 
Here we first sketch the method in section \ref{floquet}, 
and then apply it to the Falicov-Kimball 
model (one of the simplest model of correlated electrons that can be 
solved exactly within DMFT) in section \ref{fk}. From the obtained 
results, we discuss how the 
Mott-like insulator-to-metal transition is induced by AC fields 
in section \ref{discussion}.

\section{Floquet-matrix method combined with DMFT}
\label{floquet}

We outline how we can incorporate the Floquet-matrix method 
in the DMFT. Details of the method are described 
elsewhere \cite{Tsuji1,Tsuji2}. 
Our argument is based on the Keldysh Green's function formalism 
for nonequilibrium states. 
In general, a Green's function, $G(t, t')$, for a system out of 
equilibrium should depend on two time arguments 
$t$ and $t'$ independently (while in equilibrium it 
depends only on $t-t'$).  
The original idea of the Floquet method is that, 
when the driving field is periodic in 
time with the period $\tau=2\pi/\Omega$, 
there is a time-analog of Bloch's theorem for spatially periodic 
potentials.  One consequence is that 
$G$ satisfies a periodicity 
condition $G(t+\tau, t'+\tau)=G(t, t')$. 
Then one can perform the Wigner 
transformation on $G$, 
\begin{equation}
  G_n(\omega)
	  =
		  \int_{-\infty}^{\infty} dt_{\rm rel}\; \frac{1}{\tau} \int_{-\tau/2}^{\tau/2} dt_{\rm ave}\;
			e^{i\omega t_{\rm rel}+in\Omega t_{\rm ave}} G(t, t'),
\end{equation}
where $t_{\rm rel}=t-t'$ and $t_{\rm ave}=(t+t')/2$. With $G_n(\omega)$ 
we define the Floquet-matrix form of $G$ as 
\begin{equation}
  G_{mn}(\omega)
	  :=
		  G_{m-n}\left(\omega+\frac{m+n}{2}\Omega\right). 
	\label{matrix}
\end{equation}
For $G_{mn}(\omega)$ we use the reduced zone scheme, {\it i.e.}, the range of $\omega$ 
is restricted to the `Brillouin zone' on the frequency axis, $-\Omega/2 
< \omega \le \Omega/2$. With this setting $G_n(\omega)$ has a one-to-one 
correspondence to $G_{mn}(\omega)$. An advantage of employing the form (\ref{matrix})
is that the convolution of two functions $\int dt' A(t, t') B(t', t'')$ 
is mapped to the product of two matrices $\sum_{n'} A_{nn'}(\omega) B_{n'n''}(\omega)$. 

We can then employ the definition (\ref{matrix}) 
to rewrite all the self-consistent 
equations in DMFT in a Floquet-matrix form, which can be solved by the 
numerical iteration procedure as in the equilibrium case. 
The fact that we can take advantage of taking the inverse of a 
matrix when solving the Dyson equation makes our calculation numerically quite efficient. 

As an input for DMFT we need the inverse of the retarded Green's function for 
noninteracting electrons, which is given by 
\begin{align}
  ({G_{\boldsymbol k}^{R0}}^{-1})_{mn}(\omega)
	  &=
		  (\omega + n\Omega + \mu + i\eta) \delta_{mn} - 
			(\epsilon_{\boldsymbol k})_{mn},
	\label{sc_ac}
\end{align}
where 
$\mu$ the chemical potential, $\eta = 0^+$, 
and ${\boldsymbol A}(t)$ the vector potential. 
We assume that $e{\boldsymbol A}(t) = 
eE\sin(\Omega t)/\Omega\; [1,1,\dots, 1]$ 
(representing a homogeneous AC electric field in $d$ dimension).
In the hypercubic lattice, 
$(\epsilon_{\boldsymbol k})_{mn}=\epsilon_{\boldsymbol k} J_{m-n}(eE/\Omega)\;
(m-n{\rm : even})$, or $i\bar{\epsilon}_{\boldsymbol k}J_{m-n}(eE/\Omega)\; (m-n
{\rm : odd})$, where $J_n$ is the Bessel function, 
$\epsilon_{\boldsymbol k} = - 1/\sqrt{d} \sum_i \cos k_i$, and 
$\bar{\epsilon}_{\boldsymbol k}=-1/\sqrt{d}\sum_i 
\sin k_i$. An integral over $\boldsymbol k$ is calculated with the joint 
density of states $\rho(\epsilon, \bar{\epsilon})= e^{-\epsilon^2-\bar{\epsilon}^2}/\pi$
\cite{Freericks}. 

\section{Falicov-Kimball model driven by AC electric fields}
\label{fk}

Having constructed the theoretical framework, we are now in 
position to demonstrate its application to the Falicov-Kimball 
model with a heat bath. The Hamiltonian is 
$H_{\rm tot}=H_{\rm sys} + H_{\rm sys-bath} + H_{\rm bath}$ with
\begin{equation}
  H_{\rm sys}
	  =
		  \sum_{\boldsymbol k} \epsilon_{{\boldsymbol k}-e{\boldsymbol A}(t)}
			c_{\boldsymbol k}^\dagger c_{\boldsymbol k}
			+ U \sum_i c_i^\dagger c_i f_i^\dagger f_i,
\end{equation}
where $c_i^\dagger$ creates an itinerant, spinless electron while 
$f_i^\dagger$ creates a localized one, 
$H_{\rm bath}$ represents the bath, and $H_{\rm sys-bath}$ the 
coupling between the system and the bath.
We take a noninteracting 
fermion bath attached to each lattice site in the system.  
After the bath degrees of freedom are integrated out, the correction to the 
self-energy $\Gamma(\omega)$ 
is a constant, $\Gamma$, 
independent of $\omega$ in the wide band limit. 
Assuming that the bath is in 
equilibrium with a temperature $T$, we can define the nonequilibrium 
steady state (parametrized uniquely by $\Gamma$ and $T$). 
Note that the steady state is determined without 
the noninteracting Keldysh Green's function $G^{K0}$.  
Namely, the information on 
the initial state of the system and the way to switch on the field, 
on which $G^{K0}$ depends, 
is wiped out due to the dissipation caused by the bath. 

\begin{figure}[t]
  \begin{center}
	  \includegraphics[width=14cm]{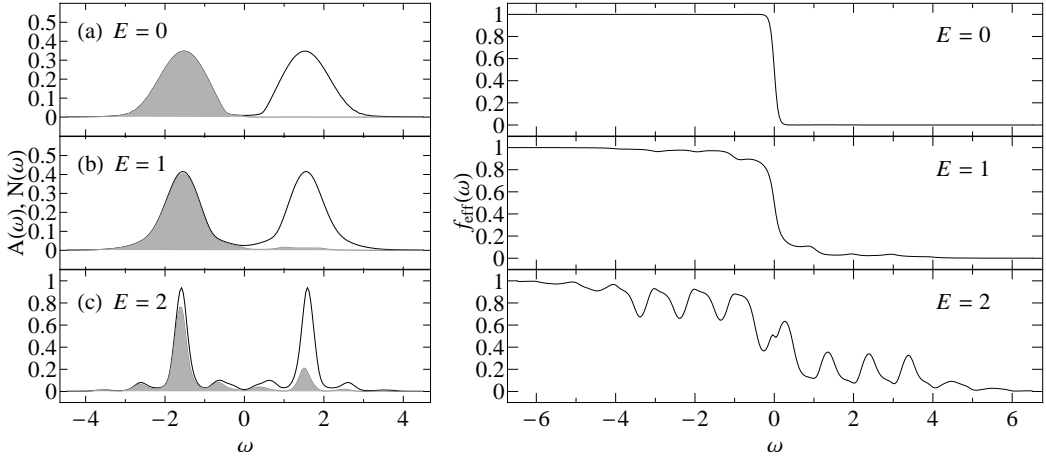}
		\caption{The local spectral function $A_0(\omega)$ 
		(solid curves) and 
    the particle number density $N_0(\omega)$ (shaded regions) 
    (left panels), and 
     the effective distribution function $f_{\rm eff}(\omega)$ (right) 
    for $E=0$ (a), $E=1.0$ (b), $E=2.0$ (c) 
    with $\Omega=1.0$, $U=3.0$, $\Gamma=0.05$ and $T=0.05$.}
		\label{ac_spec}
	\end{center}
\end{figure}
\begin{figure}[t]
  \begin{center}
	  \includegraphics[width=14cm]{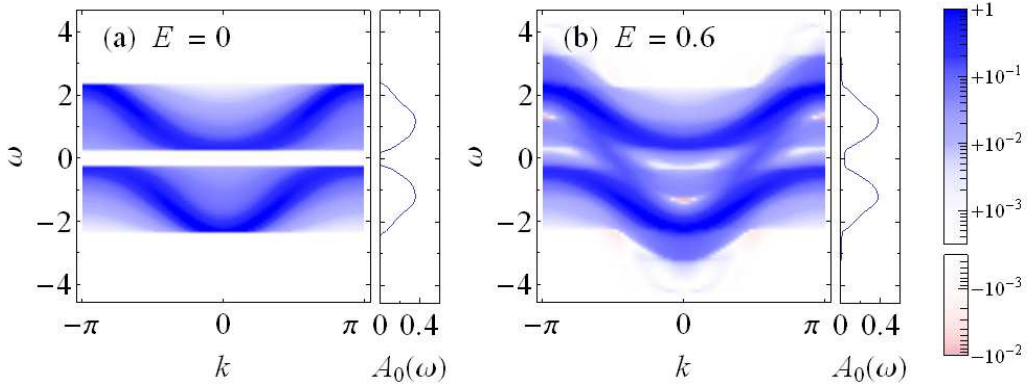}
		\caption{The gauge invariant spectral function $\tilde{A}_0({\boldsymbol k}, \omega)$
		(color-coded), and the local spectral function $A_0(\omega)$ 
		(curves) for $E=0$ (a) and $E=0.6$ (b) with $\Omega=1.0$ and $U=2.2$ 
    on the three dimensional cubic lattice, 
    and ${\boldsymbol k}=k(1,1,1)$. }
		\label{gauge}
	\end{center}
\end{figure}
We consider the following quantities: the spectral 
function $A_n(\omega) = - \sum_{\boldsymbol k} {\rm Im}
[(G_{\boldsymbol k}^R)_n(\omega)]/\pi$ (Fig.\ref{ac_spec}, left panels), the particle number density 
$N_n(\omega)=-i\sum_{\boldsymbol k}(G_{\boldsymbol k}^<)_n(\omega)/2\pi$ 
(Fig.\ref{ac_spec}), 
the effective distribution function $f_{\rm eff}(\omega)=N_0(\omega)/A_0(\omega)$
(Fig.\ref{ac_spec}, right),
the gauge invariant spectral function $\tilde{A}_n({\boldsymbol k}, 
\omega)=-{\rm Im}[(\tilde{G}_{\boldsymbol k}^R)_n(\omega)]/\pi$ (Fig.\ref{gauge}), 
and the current $j(t)=ie\sum_{\boldsymbol k} 
\bar{\epsilon}_{{\boldsymbol k}-e{\boldsymbol A}(t)} G_{\boldsymbol 
k}^<(t,t)$ (Fig.\ref{current}). Here
$G^R$ and $G^<$ are the retarded and the lesser Green function respectively.
Note that, to make the ${\boldsymbol 
k}$-dependent Green's function gauge invariant, we use the modified function
$\tilde{G}(t, {\boldsymbol r}; t',{\boldsymbol r}') = 
\exp(ie\int_{(t, {\boldsymbol r})}^{(t', {\boldsymbol r}')}dx^\nu 
A_\nu(x)) G(t, {\boldsymbol r}; t', {\boldsymbol r}')$ \cite{Boulware}.

\section{Discussion}
\label{discussion}

\begin{figure}[t]
  \begin{center}
	  \includegraphics[width=8cm]{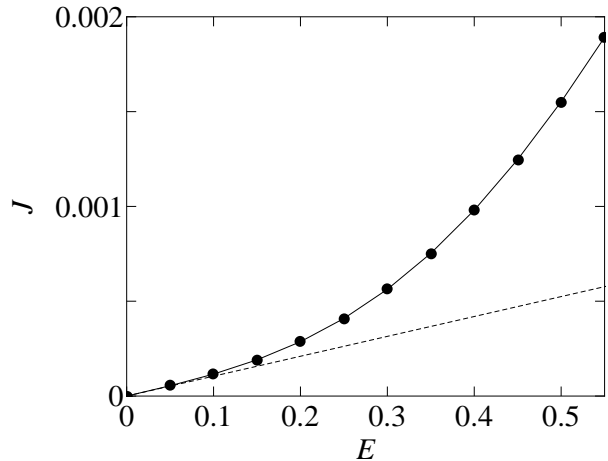}
		\caption{The current-voltage characteristic (solid curve) with $\Omega=1.0, U=3.0, 
    \Gamma=0.05$, and $T=0.05$. We also show the 
    linear-response result (dashed line) for comparison.}
		\label{current}
	\end{center}
\end{figure}
From the results one can see how the Mott insulating state is driven 
into a metallic state as one increases the intensity of the AC field 
beyond the linear-response regime. 
Specifically, we find in Fig.\ref{ac_spec} that the Mott gap collapses, 
and a new spectral weight emerges in the midgap region. Since the optical gap 
$\sim U(=3.0$ here) is greater than the frequency $\Omega(=1.0$ here), 
the direct interband transition is forbidden. 
However, if the intensity 
$E$ is sufficiently large, the electrons are excited to the conduction band 
through the midgap state, and the distribution function $f_{\rm 
eff}(\omega)$ (Fig.\ref{ac_spec}) considerably deviates from the Fermi distribution $f(\omega)=1/(e^{\omega/T}+1)$ 
realized in equilibrium. More surprisingly, 
there is a region around certain $\omega$ where 
the population exceed those in lower-energy regions, 
a kind of {\it population inversion}. 
In the band dispersion of the system out of equilibrium shown in 
Fig.\ref{gauge}, 
we again observe that the midgap state is induced by the field, whose band 
structure is roughly replicas of the original one with the spacing $\Omega$.
Finally in Fig.\ref{current}, the nonlinearity with the AC field is 
captured most clearly as the I-V characteristics 
obtained in the present method. 
Here $J \equiv {\rm Re}\; \frac{1}{\tau}\int_0^\tau dt\; e^{i\Omega t}j(t)$ 
is the current. 
When $E$ is small, the current is 
approximately proportional to $E$ 
as in the linear-response theory, but 
a nonlinear response becomes evident for $E\gtrsim 0.2$. This 
suggests that in the heart of the photoinduced insulator-metal 
transition lies the nonlinearity of the driving field. 

During the preparation of this paper, we notice that a similar idea of using a matrix 
form was mentioned in ref. \cite{Joura}.


\section*{References}
\begin{thebibliography}{9}
  \bibitem{Miyano} 
	  K. Miyano, T. Tanaka, Y. Tomioka, and Y. Tokura, 
    Phys. Rev. Lett. {\bf 78}, 4257 (1997). 
	\bibitem{Fiebig} 
	  M. Fiebig, K. Miyano, Y. Tomioka, and Y. Tokura, 
	  Science {\bf 280}, 1925 (1998). 
	\bibitem{Matsubara}
	  M. Matsubara, Y. Okimoto, T. Ogasawara, Y. Tomioka, H. Okamoto, and Y. Tokura, 
		Phys. Rev. Lett. {\bf 90}, 207401 (2007). 
	\bibitem{Shirley} 
	  J. H. Shirley, 
		Phys. Rev. {\bf 138}, B979 (1965). 
	\bibitem{Sambe} 
	  H. Sambe, 
		Phys. Rev. A {\bf 7}, 2203 (1973). 
	\bibitem{Freericks} 
	  J. K. Freericks, V. M. Turkowski, and V. Zlati\'{c}, 
    Phys. Rev. Lett. {\bf 97}, 266408 (2006). 
	\bibitem{Tsuji1}
	  N. Tsuji, 
		Master thesis, University of Tokyo, 2008.
	\bibitem{Tsuji2} 
	  N. Tsuji, T. Oka, and H. Aoki, 
		in preparation.
	\bibitem{Boulware}
	  D. G. Boulware, 
		Phys. Rev. {\bf 151}, 1024 (1966).
	\bibitem{Joura}
	  A. V. Joura, J. K. Freericks, and Th. Pruschke, 
		cond-mat/0804.3077. 
\end{thebibliography}
\end{document}